# HADOOP MAPREDUCE PERFORMANCE ENHANCEMENT USING IN-NODE COMBINERS

Woo-Hyun Lee[1], Hee-Gook Jun[2], and Hyoung-Joo Kim[3]

School of Computer Science and Engineering Seoul National University, South Korea

## ABSTRACT

*While advanced analysis of large dataset is in high demand, data sizes have surpassed capabilities of conventional software and hardware. Hadoop framework distributes large datasets over multiple commodity servers and performs parallel computations. We discuss the I/O bottlenecks of Hadoop framework and propose methods for enhancing I/O performance. A proven approach is to cache data to maximize memory-locality of all map tasks. We introduce an approach to optimize I/O, the in-node combining design which extends the traditional combiner to a node level. The in-node combiner reduces the total number of intermediate results and curtail network traffic between mappers and reducers.*

## 1. INTRODUCTION

Hadoop is an open source framework that provides a reliable storing of large data collections over multiple commodity servers and parallel processing of data analysis. Since its emergence, it has firmly maintained its position as de facto standard for analyzing large datasets. Without in-depth understandings of complex concepts of a distributed system, developers can take advantages of Hadoop APIs for an efficient management and processing of the big data.

Hadoop MapReduce [1] is a software framework built on top of Hadoop used for processing large data collections in parallel on Hadoop clusters. The underlying algorithm of MapReduce is based on a common map and reduce programming model widely used in functional programming. It is particularly suitable for parallel processing as each map or reduce task operates independent of one another. MapReduce jobs are mostly I/O-bound as 70% of a single job is found to be I/O-intensive tasks [2]. A typical MapReduce job is divided into three sequential I/O-bound phases:

(1) Map phase: Locations of input data blocks distributed over multiple data nodes are retrieved via NameNode. Blocks are loaded into memory from local disk and each map task processes corresponding blocks. Intermediate results from each map task are materialized in map output buffers. When the contents of a buffer reach a certain threshold size, they are spilled to local disk.

(2) Shuffle phase: Once a map task is completed, spilled contents are merged and shuffled across the network to corresponding reduce tasks.

(3) Reduce phase: Each reduce task process received key groups. Similar to the map phase, reduce inputs are temporarily stored in reducer output buffers and periodically spilled to disks. Once all groups are processed, final results are written to HDFS as raw files.

An increase in demand for non-batch and real-time processing using Hadoop has made performance the key issue for many MapReduce applications. A tolerable job completion time is vital for any performance-oriented jobs thus an efficient MapReduce job must aim to minimize the number of I/O operations performed in each I/O-intensive phase described above [2]. In this paper, we show how caching the input data and locally aggregating intermediate results using the in-node combiner can optimize the overall performance of a MapReduce job.

This paper is organized as follows. Section 3 gives an overview of the Hadoop MapReduce framework, describes the two bottlenecks found in MapReduce jobs and proposes two solutions for eliminating them. The algorithm for the in-node combiner, an enhancement to the traditional combiner, is demonstrated using a word count example in Section 4. Section 5 discusses the experimental results for counting daily word occurrences in Twitter messages using three different combining design patterns.

## 2. RELATED WORK

### 2.1 Hadoop Distributed File System

Hadoop Distributed File System (HDFS) [3] is a Java-based file system that provides a scalable and reliable data storage system. It is built on top of the local file system and is able to support up to few petabytes of large dataset to be distributed across clusters of commodity servers. HDFS is the basis for most of Hadoop applications. It consists of a single NameNode and a number of DataNodes. The NameNode is responsible for managing the cluster metadata and the DataNode stores data blocks. All data stored in HDFS is broken down into multiple splits and distributed throughout the DataNodes. This allows large datasets beyond a capacity of a single node to be stored economically and also enables tasks to be executed on smaller subsets of large data sets. HDFS makes several replicas (3 by default) of all data blocks and stores them in a set of DataNodes in order to prevent data lose in case of hardware failures. At least one copy is stored at a different rack and thus both fault tolerance and high availability are assured. This feature allows a cluster to operate normally even with a node failure since data is guaranteed to be stored across multiple DataNodes [4-6].

A Hadoop job is commonly divided into a number of tasks running in parallel. Hadoop attempts to schedule a task with a consideration of data block locations. It aims to allocate tasks to run at where the corresponding data block resides. This feature minimizes unnecessary data transfer between nodes.

### 2.2 Hadoop MapReduce

MapReduce [3] is one of many programming models available for processing large data sets in Hadoop. While Hadoop framework efficiently maintains task parallelization, job scheduling, resource allocation and data distribution in the backend, the MapReduce framework simply has two major components, a mapper and a reducer, for data analysis.

A mapper maps every key/value record in the dataset by arbitrary intermediate keys and a reducer generates final key/value pairs by applying computations on the aggregated pairs. The strength of MapReduce framework lies in running such simple but powerful functions with Hadoop's automatic parallelization, distribution of large-scale computations and fault tolerance features using commodity hardware.

The top-level unit of each MapReduce task is a job. A job has several mappers and reducers allocated by the underlying scheduler depending on various factors including the size of input and available physical resources. The developer, with a minimum knowledge of a distributed system, simply needs to write Map and Reduce functions which are available as Hadoop APIs in various programming languages, to take advantage of the framework. The MapReduce model can be applied to various applications including distributed grep, graph problems, inverted index and distributed sort. Figure 1 describes a workflow of a common MapReduce job.

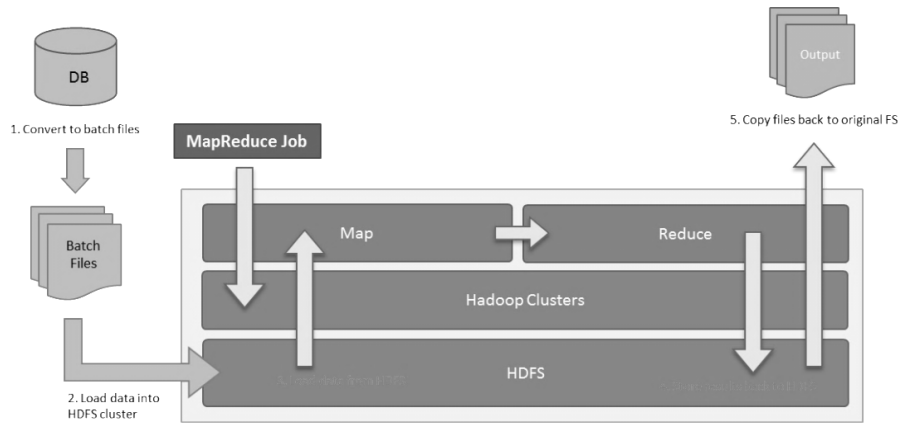

Figure 1. A workflow of typical MapReduce job.

A detailed walkthrough of a MapReduce application is now described. The Input data files stored in HDFS are split into M pieces of typically 64MB per piece and distributed across the cluster. Once a MapReduce job is submitted to the Hadoop system, several map and reduce tasks are generated and each idle container is assigned either a map task or a reduce task. A container who is assigned a map task loads the contents of the corresponding input split and invokes MAP method once for each record. Optionally on the user's request, SETUP and CLOSE methods may run prior to the first or after the last MAP method call respectively. Upon each MAP method call, it passes key and value variables to EMIT method, which then pairs are temporarily stored in a circular in-memory output buffer along with corresponding metadata. Figure 2 describes a structure of a circular map output buffer. Once the contents of a buffer reaches certain threshold size (80% by default), all key/value pairs are partitioned based on their keys and finally spilled to local disk as a single spill file per buffer. The number of partitions is equal to the total number of reduce tasks allocated for the job. Combiners, which are mini reduce tasks that combine intermediate results, may occasionally run on each partition prior to disk spills. Once all records have been processed, spill files of a task are merged as a single partitioned output file. Then each partition is transferred to the corresponding reducer across the network. This stage of the task is referred to as the shuffle phase. Figure 3 describes a workflow of the shuffle phase.

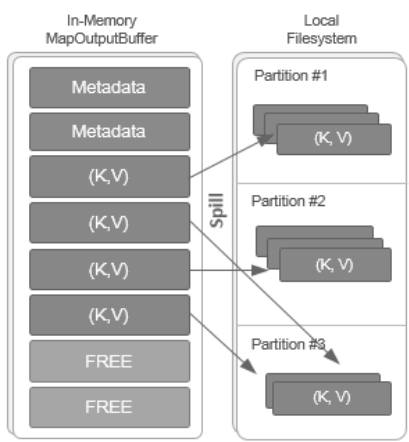 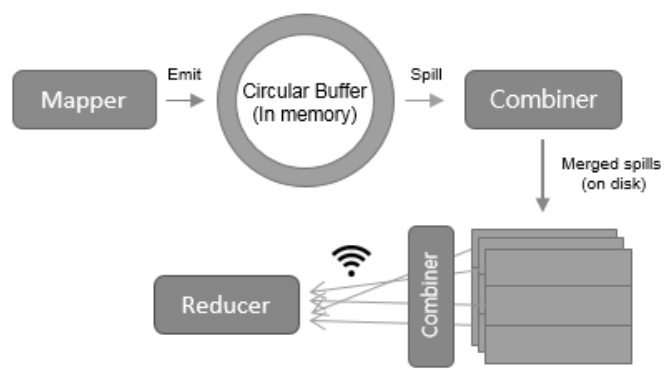

Figure 2. Circular map output buffer.   Figure 3. MapReduce shuffle phase.

The reduce task sorts and groups received intermediate pairs by their keys preferably in memory but if their sizes exceed the memory limit, an external sort is used. Once pairs are sorted, REDUCE method is invoked once per each key group and the output is appended to a final output file. Finally one output file per reduce task is stored in HDFS. Figure 4 describes an example of a MapReduce job.

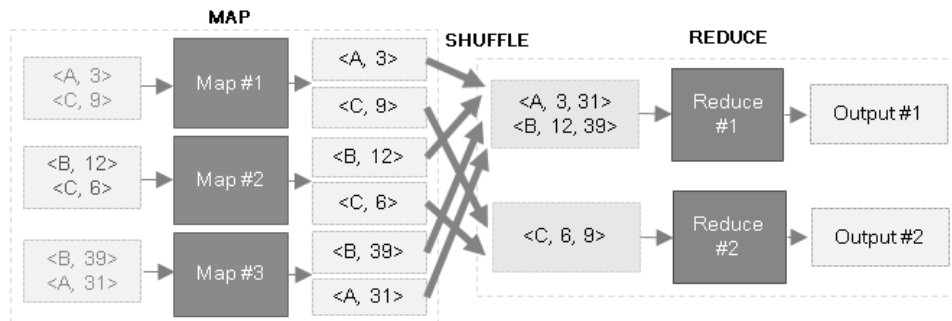

Figure 4. An example of a MapReduce job.

## 2.3 Hadoop I/O optimization

The most mentioned weakness of HDFS is its poor I/O performance. Attempts to solve this problem can be classified into either combining stored files into forms of databases or modifying the existing HDFS I/O features [7]. The former approach improves system throughput rather than I/O performance by providing efficient indexing of data blocks. The second approach requires a complete re-design of the entire Hadoop system, which comparatively is dangerous. As a simple but practical alternative, utilizing an in-memory data storage system to cache input data is proven to be the most effective method for improving I/O performance of any data-intensive tasks.

Ananthanarayanan et al. [2] built PACMan, an input data caching service that coordinates access to the distributed caches. Two cache eviction policies, LIFE and SIFE, are implemented within PACMan. LIFE evicts the cached blocks of the largest incomplete file and SIFE replaces cached blocks with the smallest incomplete file. These two policies aim to optimize for job completion time by maximizing memory-locality of tasks. Overall job completion times were reduced by up to 53% with LIFE and cluster utilization improved by up to 52% with SIFE.

Zhang et al. [7] pointed out the poor HDFS file access performance as the major drawback of Hadoop. In order to provide high access performance without altering the existing HDFS I/O features, they built a novel distributed cache system named HDCache which periodically makes snapshots of local disk in shared in-memory caches that are forged as local disks to Hadoop. By storing replicas in different caches for every cached files, disk I/O is substituted for either local memory access or network I/O which leads to a significant improvement in overall performance.

Senthikumar et al. [8] implemented Hadoop R-Caching, a caching system that adopts an open source in-memory database, Redis, as both global and local cache layers for HDFS. Redis, a high performance in-memory key-value storage, has been proven for its stability and efficiency not only as a database but also as a cache for Hadoop.

While caching input data to maximize memory-locality of MapReduce tasks significantly reduces disk I/O operations in the map phase, I/O bottleneck during the shuffle phase is a significant performance degradation factor. Crume et al. [9] showed preliminary designs of approaches to compress intermediate data, which up to five orders of magnitude reduction the original key/value ratio was observed.

Dean and Ghemawat [1] suggested using combiners to reduce the size of intermediate results in MapReduce jobs. Lin and Schatz [10] introduced the in-mapper combining design, which is an improvement of the traditional combiner. This design guarantees the execution of combiners by moving the combining function within the map method.

## 2.4 NoSQL

While many modern applications require data with various formats and sizes to be stored and accessed simultaneously, typical Relational databases do not meet these requirements as they are not optimized for scalability and agility challenges. Most relational databases require data schema to be strictly defined and guarantee ACID properties to ensure database reliability. ACID properties are:

- Atomicity: Each transaction is atomic that either a transaction is fully completed or not executed at all. A failure of a part of transaction must lead to a failure of an entire transaction.

- Consistency: Only valid information is written to the database. All operations must abide by customary rules and constraints.

- Isolation: Each transaction is isolated from any other transactions running concurrently. Concurrent transactions must not interfere with each other.

- Durability: Committed transactions must be stored permanently even in the event of system failures or errors. Restoration of committed transactions should be ensured through database backups and transaction logs.

ACID properties guarantee the database reliability but their strictness are not suitable for simplicity and scalability which many modern applications require. NoSQL (Not Only SQL) database [11] is developed in response to a rise in volume of data and high data access/write performance. NoSQL databases generally do not require a predefined schema thus data with various formats can be easily added to the application without significant changes. In oppose to ACID properties, NoSQL databases are based on the BASE paradigm and the CAP theorem. BASE paradigm stands for Basically Available, Soft state, Eventually consistency. It makes a tradeoff to consistency for availability and performance. NoSQL databases can achieve only two of the three CAP theorem. Either a system guarantees consistency and partition tolerance, availability and partition tolerance or consistency and availability. By sacrificing some strengths relational databases have, NoSQL database is able to provide highly scalable system which large volume of data is distributed across commodity servers and thus high read/write performance is achieved [12-14]. There are four types of data models supported by NoSQL databases:

- Key-value: records are stored as key-value pairs. (Redis, Memcached, Dynamo)

- Column oriented: records are stored as sparse tabular data. (Bigtable, Cassandra, HBase)

- Document oriented: each record is a document that contains multiple fields. (MongoDB)

- Graph oriented: records are stored as graph nodes and edges. (Neo4j)

Support for flexible data models and high performance make NoSQL database a perfect choice for caching frequently accessed/modified data. NoSQL databases have been adopted as both dynamic caches and primary data stores by various enterprises. In this paper, we utilize Redis, an in-memory NoSQL database, as our cache layer for both input data and intermediate results of MapReduce jobs.

## 3. BACKGROUND

MapReduce framework is a powerful model for processing large datasets in a distributed environment. As described in the previous section, each MapReduce phase requires multiple disk and network I/O operations. A typical MapReduce job consumes relatively low resource on computing whereas 79% of a job is I/O intensive [2]. In order to improve overall performance of a MapReduce job, unnecessary I/O operations must be minimized. In this section, we identify two significant I/O bottlenecks faced by MapReduce jobs and solutions for resolving those issues.

### 3.1 HDFS bottleneck

The poor performance of Hadoop is rooted in its nature of batch processing and HDFS, which is optimized for high throughput rather than high I/O performance. Redesigning the processing module can solve the former cause. However the weakness of HDFS inherently is caused by underlying hardware and its design principles [15].

HDFS is primarily designed for storing and processing vast volumes of data. It follows write-once-read-many model, which thus simplifies data coherency and enables high throughput access. However, such requirement has led to a comparatively large data block size (64MB by default) and consequentially resulted in inefficient random write and read performance. Data-intensive tasks such as MapReduce jobs require high file access performance. Once a MapReduce job is submitted, NameNode retrieves locations of all data blocks needed for the job then each allocated task loads blocks from local disk to memory and processes each records. While Hadoop tries to maximize data locality by assigning tasks at nodes where the target data resides, loading multiple large blocks into memory is still significant performance degrading operations. Without modifying the core of HDFS, reducing HDFS I/O within a MapReduce job is the most effective approach for enhancing file access performance.

### 3.1.1 In-memory cache

Utilizing an in-memory cache to maximize memory-locality of a MapReduce job has been proven to be efficient for reducing HDFS I/O operations. An additional thread periodically loads data blocks stored in HDFS into in-memory cache and evict them according to appropriate eviction policies and task schedules. Instead of directly loading large data blocks from HDFS to memory at every data request, caches are queried for data availability as a priority. A significant improvement in performance is guaranteed when all input data is cached and hence HDFS I/O during the data read phase is at its minimum [2]. In-memory cache systems such as Memcached and Redis provide not only high throughput but also high file access rate and are adequate choices for caches. Figure 5 describes an overview of an in-memory cache for a MapReduce job.

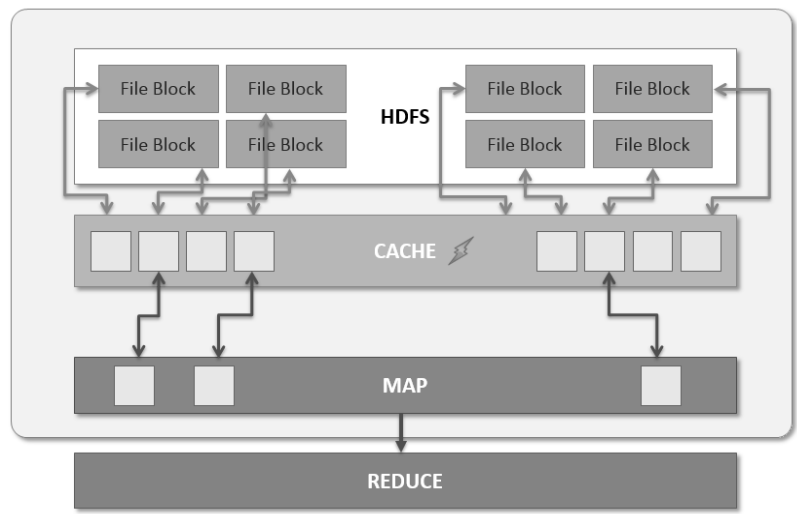

Figure 5. An in-memory cache for Hadoop MapReduce.

### 3.2 In-Memory cache

During the shuffle phase of a MapReduce job, intermediate results generated by a map task are temporarily stored in a circular output buffer and periodically spilled to disk and finally shuffled to corresponding reducers across the network. The total number of I/O operations during this phase depends on the amount of intermediate results and the number of reducers to transfer to. Reduce tasks generally do not begin reduce functions until all input data have been processed by map tasks. The time taken to process all records and transferring intermediate pairs to corresponding reducers account for significant portion of overall processing. A research [16] shows that the shuffle phase accounts for 26%-70% of the running time of 188,000 MapReduce jobs ran by Facebook. This confirms that transferring data between successive phases is a severe bottleneck in MapReduce jobs. Hence, optimizing network activity at this phase is critical for improving job performance. As the most simple but efficient solution for minimizing the volume of intermediate data emitted by map tasks, we introduce three different combining design patterns in this section.

The combiner function [3] is a useful extension provided as a Hadoop API that performs partial merging of intermediate data prior to sending them across the network to reducers. In a case where intermediate results contain significant number of repetitions that are destined for the same reducer; the combiner can substantially reduce the amount of intermediate results and therefore save substantial network communication cost without altering the final outputs.

#### 3.2.1 Combiner

The combiner is a mini-reducer that operates on data generated by map tasks. It is executed in isolation per task and performs local aggregation between map and reduce tasks to curtail network traffic. A combiner function in general is identical to the reduce function except its output types must match reducer's input types. Combiners by implementation are designed to run at most twice during the map phase. The first run is prior to spilling of contents stored in each map output buffer and the second run is on merging stage of spill files at the end of a map task. Theoretically combiners should substantially improve overall performance of MapReduce jobs with high population of combinable intermediate results by cutting down the network communication cost. However, two significant drawbacks lie within using combiners:

- Execution of a combiner is not guaranteed: Combiners may not be executed on some occasions as Hadoop may choose not to run them if execution is determined to be inefficient for the system. A known but configurable occasion is when the number of spill files does not exceed the configured threshold (3 by default). Other occasions are systemically not controllable by developers. Such randomness may cause undesired situations where combinable intermediate results are not fully combined thus missing out on potential optimizations.

- Size of emitted map outputs is not optimized: The emitted results are temporarily stored in in-memory buffers and the combining function is applied on them before spilling them to local disk. Thus combiners do not actually reduce the number emitted results. This characteristic leads to situations where map output buffers are filled with soon-to-be combined outputs causing more spill files to be generated.

### 3.2.2 In-Mapper combiner

The in-mapper combiner (IMC) [10] resolves the two problems of the traditional combiner addressed above. The key idea of IMC is to run the combining function inside the map method to minimize the volume of emitted intermediate results. Instead of emitting results to map output buffers at every invocation of the MAP method, IMC stores and aggregate results in an associative array indexed by output keys and emit them at the end of the map task. This approach guarantees the execution of combiners and substantial reduction in the total number of emitted map outputs. Figure 6 shows a pseudo code for a word count MapReduce job with IMC design pattern. The total number of map outputs sent across the network is O(R) for a simple word count MapReduce job without a combiner and O(KM) for a job with IMC, where R corresponds to the total number of input records, K corresponds to the number of distinct keys in the dataset and M corresponds to the total number of allocated mappers for the job. Because the scope of IMC is bound to a mapper and its execution is guaranteed and the effectiveness of IMC increases relative to the total number of mappers, which by far is smaller than the total number of records. Figure 7 shows an example of a MapReduce job using in-mapper combining design.

```
1: class Mapper
2:     method Setup()
3:         H ← InitAssociativeArray()
4:     method Map(long id, twit t)
5:         d ← ExtractDate(t)
6:         W ← BagOfWords(t)
7:         for all words w ∈ W do
8:             H{d, w} ← H{d, w} + 1
9:     method Cleanup()
10:        for all date-word-pair dw ∈ H do
11:            Emit(date-word-pair dw, count H{d, w})
1: class Reducer
2:     method Reduce(date-word-pair dw, counts [c1, c2, ....])
3:         s ← InitCount()
4:         for all count c ∈ counts do
5:             s ← s + c
6:         Emit(date-word-pair dw, sum s)
```

Figure 6. Algorithm 1: Word count algorithm with IMC design pattern.

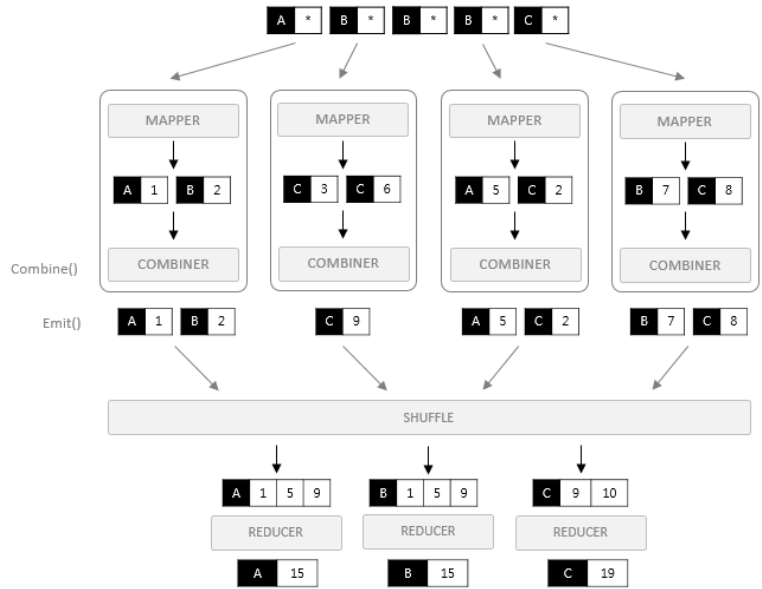

Figure 7. A MapReduce job with the in-mapper combining design.

## 4. OUR APPROACH

The in-mapper combiner is capable of resolving the problems of traditional combiner and improves the overall performance substantially. The combining function of a traditional combiner runs in a separate thread from the main mapper thread. As long as the map output buffer is not fully occupied, the map method is executed in parallel with the combining function. However, in order to guarantee execution of combiners, IMC withdraws parallelism by moving the combing function within the map method. Each map task is required to maintain an associative array for storing intermediate results. Often when dealing with large data sets with IMC, the size of distinct keys stored in an associate array exceeds heap size of a map task therefore causing a memory overflow. An explicit memory management is necessary for such case. When the size of the array grows beyond its capacity at key insertion, least recently updated records are evicted and emitted to buffers to free up memory.

Similar to the traditional combiner, the scope of IMC is limited within a single map task. However Hadoop's strength lies in its capability for parallel processing. Typically multiple map tasks each processing different data splits run in each node in parallel. Taking this into account, the scope of IMC can be extended to a node-level by combining all intermediate results generated within the same node for further optimization. As an improvement to IMC, we propose a new combing design pattern called the in-node combiner.

### 4.1 In-Node Combiner (INC)

The key idea of the in-node combiner is to combine all intermediate results generated within a node. Instead of maintaining a single associative array for each map task, arrays are merged into a single locally shared data structure that stores all intermediate results in the same node. All map tasks aggregate results in a locally shared cache and the last map task running in the node emits results stored in the cache. Figure 8 shows a pseudo code for a word count MapReduce job with the in-node combining design pattern.

The in-node combiner has two significant benefits over IMC:

International Journal of Computer Science & Information Technology (IJCSIT) Vol 7, No 5, October 2015

- Total number of emitted results by a node is minimized: The domain of local aggregation is extended to node level leading to a further reduction in the total number of emitted intermediate results. By consuming smaller portion of map output buffers and forcing only the last mapper to emit locally combined results, fewer spill files are generated. Finally, reduced intermediate result size guarantees substantial reduction in network communication cost.

- Combining function is executed in a separate thread: IMC made a tradeoff between parallelism and performance. Combining function was replaced into the map method. However, by using an in-memory cache system that runs outside of Hadoop for storing intermediate results, INC shifts the responsibility for combining, managing memory and indexing to a separate thread.

```
1: class Mapper
2:      method Setup()
3:          C <- InitCache()
4:      method Map(long id, twit t)
5:          d ← ExtractDate(t)
6:          W ← BagOfWords(t)
7:          for all words w ∈ W do
8:              C{d, w} ← C{d, w} + 1
9:      method Cleanup()
10:         for all date-words dw ∈ H do
11:             if ( C{d, w} > threshold OR isLastMapper )
12:                 Emit( {d, w}, count C{d, w} )
```

Figure 8. Algorithm 2: Word count algorithm with INC design pattern.

In order to prevent cache overflows due to excessive amount of distinct keys, two properties are checked at map method invocation. If a key has a value larger than a certain threshold (pre-emit threshold), it is immediately emitted by the current map task. The number of results emitted by the last mapper thus is slightly reduced. This approach is particularly effective for partially sorted initial data sets where similar keys are likely to be handled by the same map task. A task also periodically checks the current cache size and evicts a portion of combined results to free up memory.

The number of intermediate results transferred across the network decreases to $O(KN)$ for a word count MapReduce job using INC, where N corresponds to the total number of data nodes. The performance of INC increases relative to the number of data nodes in the cluster. When pre-emit threshold is set to infinity and memory is sufficient enough to store all keys, the total number of network I/O operations is equal to the sum of distinct keys stored in each node cache. The number of participating data nodes by principle is far smaller than the number of allocated map tasks, thus substantial performance enhancement is expected with INC. Figure 9 and 10 describe an overview of a MapReduce job using INC.

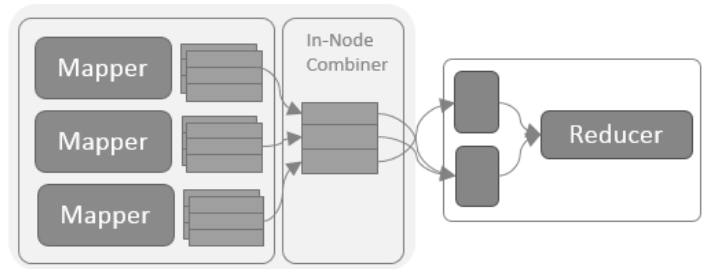

Figure 9. An overview of a MapReduce job with in-node combiner.

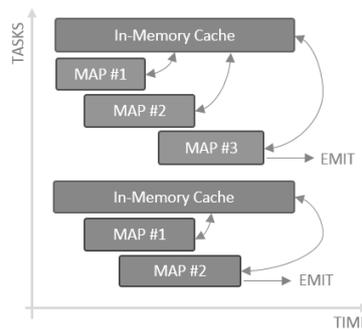

Figure 10. In-node combiner example.

## 4.2 Implementation

Our architectural goal is to avoid altering the existing Hadoop features. Modifying the core of any systems is not only complex but may violate the original design principles. Any newly implemented features must be fully compatible with and independent of existing Hadoop features. Thus aggressive use of Hadoop API and other existing stable systems are the prior considerations for our implementation.

An implementation of a Hadoop in-memory cache can be either an entirely new system designed primarily for Hadoop or a modification of an existing cache system. For our purpose, we chose to make a use of an existing cache system that best satisfies our requirements. Of many available cache systems, Memcached and Redis are the two most predominant in-memory key/value data stores available as open-source. Their usage is not only limited to caches but also primary databases for various applications [17].

REmote DIctionary Server known as Redis is an open-source in-memory data structure store. It is a popular key-value cache and a database. One notable difference Redis has compared to Memcached is that keys in Redis can be mapped to non-string data types including lists, sets, sorted sets and hashes allowing data to be stored and handled is various formats. Redis also supports full snapshot mechanism and disk serialization. Either data is asynchronously stored to disk periodically or all data modifying operations are logged in log files. Although in its beta phase, Redis also provides full clustering features that include auto partitioning, live reconfiguration and fault tolerance. Our quad-core machine can process 232k SET requests per second and 227k GET commands per second.

Redis provides a built-in protection allowing the user to set a max limit to memory usage. Redis will either return error messages to write commands or evict least recently used keys when the max memory limit is reached. Redis can handle up to 232 keys in a single instance. An empty instance used about 1MB and 1 million hashes with 5 fields occupy only around 200MB. Due to its exceptional read/write performance, support for various data types and efficient memory usage, Redis is the perfect choice for our cache and a data store.

**4.2.1 System architecture**

We set multiple Redis instances at each node, which are clustered into a single global Redis instance. Performance enhancement is guaranteed only when the entire input data for the job is cached, an occasion where only a fraction of data is cached may even lead to performance degradations. In order to observe the effects of fully loaded caches with 100% memory hit ratio, we deliberately loaded the entire data into the Redis cluster. Each record is stored in a hash with multiple fields. Hash types in Redis has a constant lookup speed.

A custom InputFormat is implemented to directly read each hash bucket from local Redis instances instead of regular batch files. The RedisHashInputFormat assigns each Redis instance as a single input split, therefore the number of allocated map tasks for a job is equal to the total number of local Redis instances. The RedisHashInputFormat retrieves a list of all keys stored in the corresponding local Redis instance at its initialization. Each record is retrieved at each nextKeyValue method invocation.

A container in Hadoop is a collection of physical resources allocated by the ResourceManager upon job submission. The number of allocated containers varies by the required resources for the submitted job. RMContainerAllocator class is responsible for allocating either map or reduce tasks to containers. In our system, upon assigning a map task to a container, each container establishes a connection to the local Redis cache and updates the total number of allocated map tasks within the same node under a configured key. Each map task also updates its status on task completion in the local cache allowing other map tasks in the same node to be aware of overall job status. Each map task compares the total number of map tasks to the completed map tasks stored in the cache to verify if it is the last mapper running in the node. Instead of storing intermediate results in an isolated associative array, they are stored in the local Redis cache. For memory efficiency, each intermediate key-value pair is stored under one of many hash buckets. Figure 11 and 12 shows an overview and a workflow of our system respectively.

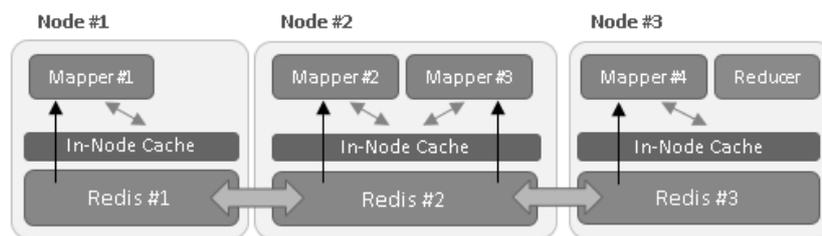

Figure 11. System overview.

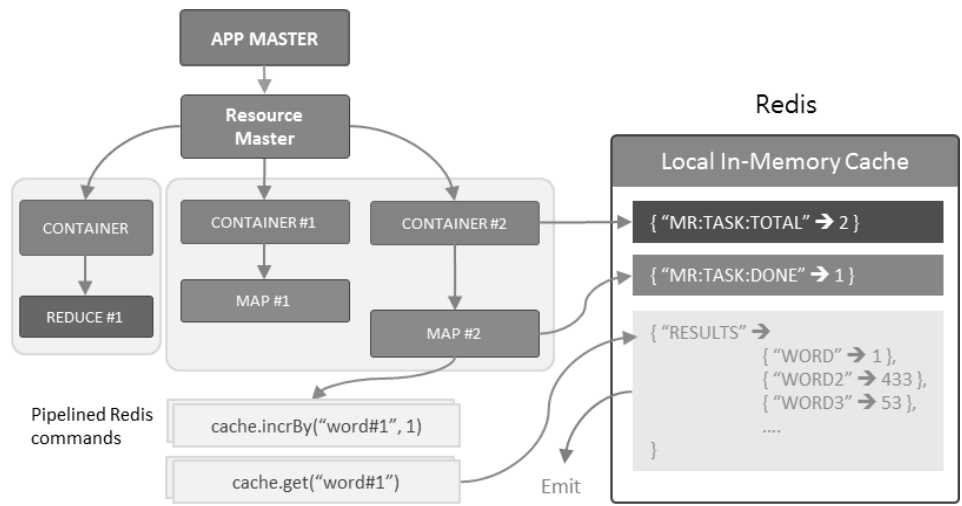

Figure 12. System workflow.

## 5. EXPERIMENT

Our Hadoop cluster consists of four physical nodes each running CentOS 6.5 with Hadoop 2.5.1 and equipped with a Intel i7 Quad-Core CPU, 8GB of RAM and 11TB HDD. Three Redis 2.9 instances run at each node, of which two are globally clustered and the other is used as a local in-node combiner cache.

The dataset used for the experiment is a set of random Twitter messages known as tweets published in March of 2013. A tweet has 6 fields; tweet id, message, original tweet id, date of submission and user id. Each tweet is separated by a new line character and multiple duplicates may exist due to the retweet feature. There are total of 20 files (12GB) each containing different number of unsorted tweets.

We implemented two simple MapReduce algorithms for performance comparisons. The main algorithm is a word count algorithm that counts occurrences of every word in Twitter messages and outputs results in separate files per day. Another MapReduce job computes relational status between Twitter users using mention tags. Twitter's mention feature directs a message to a particular user by writing username followed by at-sign. If a tweet contains mention tags, its author and the mentioned user are expected to have a relationship. Messages and referring user ids are aggregated per user using our algorithm. Figure 13 shows a pseudo-code for our second MapReduce job.

```
1: class Mapper
2:       method Map(long id, twit t)
3:           M ← getMentionTags (t)
4:           for all users m ∈ M do
5:             Emit(UserID u, MentionId, m)
6:                Emit(m, getMessage(t))

1: class Reducer
2:       method Reduce(userID u, C [c1, c2, c3, …..])
3:           for all object c ∈ C do
4:             if IsObject(c)
5:                  T ← c
6:             else
7:                T.updateStatus(c)
8:           Emit(userID u, status T)
```

Figure 13. Algorithm 3: Computing relationships between Twitter users.

## 5.1 In-Memory cache

For read performance comparisons between HDFS and an in-memory cache, all 20 files are copied into HDFS and also loaded into the Redis cluster of 8 instances (100% cache hit ratio). Each tweet is stored as a key/value pair under a hash. There are total of 8 map tasks (2 tasks per node) and each takes a single Redis instance running in the corresponding node as its input split.

As Table 1 and Figure 14 show, the average job completion time of a word count MapReduce job is reduced by 23% when using an in-memory cache. The reduction is caused by shorter map completion time which was reduced by 14%. Bypassing HDFS and using an in-memory cache as the data source substantially improves overall performance of a MapReduce job.

Table 1. Comparison between HDFS and in-memory cache.

| Data Source | Map Completion Time (min) | Job Completion Time (min) |
|---|---|---|
| HDFS | 46.53 | 68.23 |
| In-memory cache | 39.64 | 52.53 |

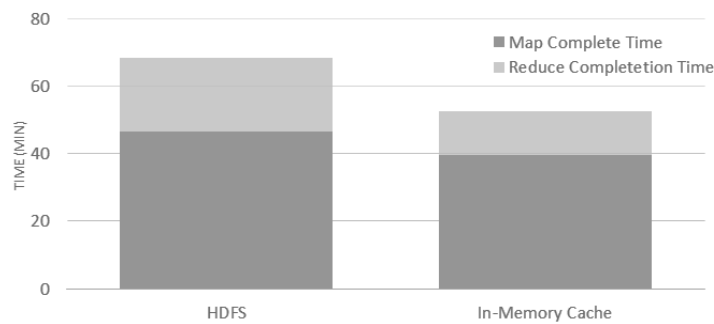

Figure 14. Average job completion time for HDFS and in-memory cache.

## 5.2 Combiner

The effects of three combining design patterns are compared for three different input sizes and cluster sizes. The same word count MapReduce algorithm with HDFS as the source of input data is used for all three combining design patterns. The number of map tasks varies by the corresponding input data size and the number of reduce tasks is fixed at one. The job completion times are for one job iteration.

Table 2. Results for different combining designs (R = 24M, N = 4)

| Method | Map Output | Reduce Input | Job Completion Time (min) |
| --- | --- | --- | --- |
| No combiner | 144,237,557 | 144,237,557 | 66.48 |
| Traditional combiner | 144,237,557 | 65,385,683 | 54.53 |
| In-mapper combiner | 65,385,683 | 65,385,683 | 48.47 |
| In-node combiner | 2,535,467 | 2,535,467 | 43.02 |

Table 2 shows results for processing 24M records with 4 datanodes. Results indicate that all three combining design patterns show significant reduction in reduce input size compared to the uncombined. Reduce input size is reduced by more than 50% and average job completion time is reduced by 30% with INC. Map output size of the traditional combiner remains unchanged from the uncombined because traditional combiners run on emitted outputs. INC generates the minimum number of map output among all combining designs. Almost 90% reduction in map output size is observed. Figure 15 shows results for a word count job with different input sizes and combining design patterns. As the input data size increases, more keys are processed by each map task and thus more pairs with a same key are combined. With INC, average job completion time was reduced by almost 50% compared to the uncombined. When the number of combinable results is large enough, INC is the most effective choice for enhancing overall performance. Results show that the effectiveness of INC increases relative to the total number of distinct keys.

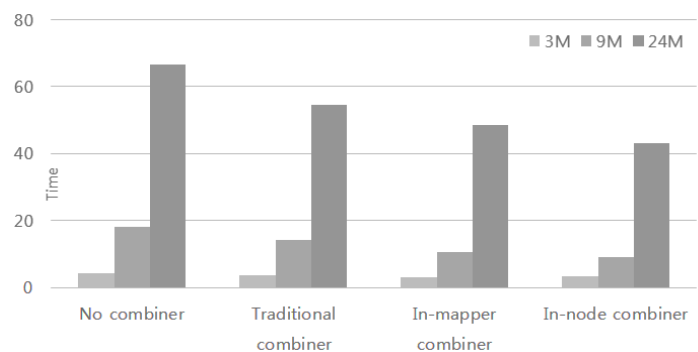

Figure 15. Average job completion time VS Input size.

Figure 16 shows results for a word count job processing 9 million records running with different cluster sizes. For all types of combining methods, increasing the cluster size improves the task parallelism and thus job completion time is greatly reduced. For a single node cluster, increase in job completion time is observed for INC due to the additional cost for maintaining connections to the local cache. However, INC performance enhances gradually with increase in cluster size. 40% enhancement in job completion time compared to the uncombined is observed with INC running in 4 data nodes.

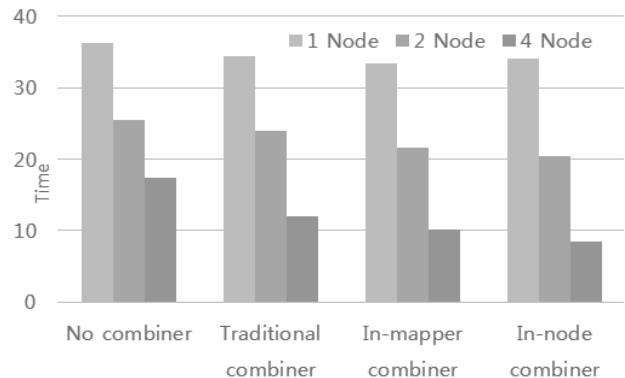

Figure 16. Average job completion time VS Number of nodes.

Unfortunately combiners do not always improve performances of all MapReduce jobs. Combiners should only be used for jobs with sufficient amount of combinable intermediate results. If the amount of intermediate pairs with a same key generated within a map task is low, using a combiner is unlikely to improve the performance but only adds additional execution costs. Table 3 shows results for running our second algorithm. Since only tweets containing mention tags are candidates for the algorithm, the number of distinct keys is significantly reduced compared to the word count algorithm. Unless messages are exactly identical to each other, each tweet with mention tags generates multiple intermediate pairs that are not combinable. Results show only 3% of total map outputs and 6% of job completion time was reduced when using combiners. This indicates that using combiners on MapReduce jobs with small percentage of combinable intermediate pairs do not have significant impact on overall performance. Combiners must be used carefully only on appropriate cases otherwise performance may be deteriorated.

Table 3. Results for computing relational status (R = 24M, N = 4)

| Method | Map Output | Reduce Input | Job Completion Time (min) |
| --- | --- | --- | --- |
| No combiner | 2,651,123 | 2,651,123 | 47.32 |
| Traditional combiner | 2,651,123 | 2,618,876 | 45.53 |
| In-mapper combiner | 2,618,876 | 2,618,876 | 44.47 |
| In-node combiner | 2,570,458 | 2,570,458 | 45.02 |

## 6. CONCLUSION

We have shown a workflow of a common Hadoop MapReduce job and described the two bottlenecks which primarily is caused by the poor I/O performance. Both disk and network I/O during all phases of a MapReduce job should be optimized for better performance. Caching entire input data of a job ensures a significant improvement in overall performance. Though caching solves the HDFS bottleneck by completely bypassing it but multiple disk and network I/O performed during the shuffle phase are significant performance degradation factors. The combiner was introduced to reduce the amount of intermediate results shuffled across the network by locally aggregating partial results at the map side. The in-mapper combiner improves the traditional combiner by reducing the number of emitted intermediate results. Our experimental results showed that the job completion time was reduced by 25% using an in-mapper combiner. The effectiveness of IMC is relative to the total number of allocated map tasks. We proposed the in-node mapper which extends the scope of IMC to node level. It aims to combine all intermediate results within the same node by locally combining intermediate results generated within the same node. Our experimental result showed INC improves the job performance by up to 20% compared to IMC.

We have modified Hadoop core to utilize in-memory cache to store intermediate results and map task status information. Our system allows map tasks to be aware of current status of the node it is running on. Using this feature, various different combining techniques can be applied to further optimize MapReduce jobs.